\documentclass[aps,prd,amsmath,amssymb,twocolumn,superscriptaddress,preprintnumbers,nofootinbib]{revtex4-1}
\pdfoutput=1
\usepackage{graphicx}
\usepackage{amsthm}
\usepackage{amsmath}
\usepackage{graphicx,subfigure}
\usepackage{dcolumn}
\usepackage{bm}
\usepackage{slashed}

\newcommand{\be}{\begin{eqnarray}}
\newcommand{\ee}{\end{eqnarray}}
\newcommand{\bea}{\begin{eqnarray}}
\newcommand{\eea}{\end{eqnarray}}

\newcommand{\keV}{{~\rm keV}}
\newcommand{\MeV}{{~\rm MeV}}
\newcommand{\GeV}{{~\rm GeV}}

\newcommand{\gstar}{g_{\star}}
\newcommand{\YN}{Y_{\rm s}}
\newcommand{\mN}{m_{\rm s}}
\newcommand{\Mpl}{M_{\rm Pl}}
\newcommand{\DeltaR}{\Delta_{_{\rm R}}}
\newcommand{\DeltaI}{\Delta_{_{\rm I}}}
\newcommand{\Zp}{Z'}
\newcommand{\MZp}{M_{_{Z'}}}
\newcommand{\gZp}{g'}

\begin{document}

\title{A Dark Matter Progenitor: Light Vector Boson Decay into (Sterile) Neutrinos}
\author{Brian Shuve}
\email{bshuve@perimeterinstitute.ca}
\affiliation{Perimeter Institute for Theoretical Physics 31 Caroline St. N, Waterloo, Ontario, Canada N2L 2Y5.}
\affiliation{Department of Physics \& Astronomy, McMaster University 1280 Main St. W. Hamilton, Ontario, Canada, L8S 4L8.}
\author{Itay Yavin}
\email{iyavin@perimeterinstitute.ca}
\affiliation{Perimeter Institute for Theoretical Physics 31 Caroline St. N, Waterloo, Ontario, Canada N2L 2Y5.}
\affiliation{Department of Physics \& Astronomy, McMaster University 1280 Main St. W. Hamilton, Ontario, Canada, L8S 4L8.}


\begin{abstract}
We show that the existence of new, light gauge interactions coupled to Standard Model (SM) neutrinos give rise to an abundance of sterile neutrinos through the sterile neutrinos' mixing with the SM. Specifically, in the mass range of MeV-GeV and coupling of $\gZp\sim 10^{-6} - 10^{-3}$, the decay of this new vector boson in the early universe produces a sufficient quantity of sterile neutrinos to account for the observed dark matter abundance. Interestingly, this can be achieved within a natural extension of the SM gauge group, such as a gauged $L_\mu-L_\tau$ number, without any tree-level coupling between the new vector boson and the sterile neutrino states. Such new leptonic interactions might also be at the origin of the well-known discrepancy associated with the anomalous magnetic moment of the muon. 
\end{abstract}

\pacs{12.60.Jv, 12.60.Cn, 12.60.Fr}
\maketitle


The vector-like nature of the SM quarks and leptons below the electroweak scale, coming in pairs of left- and right-handed fermions, naturally suggests the possibility of right-handed neutrinos complementing the known left-handed ones. Since they are entirely uncharged under the SM gauge groups, these right-handed neutrinos are often referred to as sterile neutrinos. More importantly, from the point of view of quantum field theory, their neutrality under the SM precludes any knowledge of their mass scale from theoretical considerations alone. However, it was recognized some time ago that they may serve as good Dark Matter (DM) candidates \cite{Olive:1981ak} for sterile neutrino masses in the range $1 - 100\keV$~\cite{Dodelson:1993je,Dolgov:2000ew,Wu:2009yr}. The relic abundance of sterile neutrinos can be produced by the SM electroweak interactions via the sterile neutrinos' mixing with left-handed neutrinos. Current constraints from X-ray data~\cite{Abazajian:2001vt,Watson:2006qb,Boyarsky:2007ay,Watson:2011dw,Horiuchi:2013noa} and small-scale structure formation~\cite{Abazajian:2005xn,Seljak:2006qw,Viel:2006kd,Boyarsky:2008xj,Polisensky:2010rw,deSouza:2013wsa,Viel:2013fqw,Horiuchi:2013noa} exclude the original Dodelson-Widrow proposal of sterile neutrino production through SM electroweak interactions as the principal mechanism for generating the DM abundance seen in cosmological data~\cite{Ade:2013zuv}, although resonant production of sterile neutrinos in the presence of a lepton asymmetry~\cite{Shi:1998km,Abazajian:2001nj,Laine:2008pg}, production through new singlet states coupling to sterile neutrinos~\cite{Shaposhnikov:2006xi,Kusenko:2006rh,Petraki:2007gq,Khalil:2008kp,Merle:2013wta}, or dilution of thermal sterile neutrino abundances through entropy production \cite{Scherrer:1984fd,Asaka:2006ek,Bezrukov:2009th,Nemevsek:2012cd} remain viable possibilities (see ref.~\cite{Kusenko:2009up} for a comprehensive review).

Quite independent of the above considerations, and motivated by a variety of experimental and observational anomalies, recent years have seen significant research into the possibility of a new force, with a mass of the corresponding gauge boson in the range of $\MeV - \GeV$, and which is feebly coupled to the SM~(e.g. \cite{Finkbeiner:2007kk,Pospelov:2007mp,ArkaniHamed:2008qn,Essig:2013lka}). While the details differ for each such model, many extensions share the exciting possibility of directly producing the new states in on-going as well as planned experiments~(e.g.~\cite{ArkaniHamed:2008qp,Bjorken:2009mm,Essig:2013lka}. Considerable attention was given to models with new vector bosons that are kinetically mixed with the gauge-boson of hypercharge, or models with new scalars that mix with the Higgs boson. These result in a coupling of SM particles to the new gauge interaction that is proportional to either the particle's electromagnetic charge or to its mass.  In either case, the coupling to neutrinos does not play a significant role. In this \emph{Letter}, we consider instead a new light vector boson that couples primarily, albeit weakly, to SM leptons, including the left-handed neutrinos. It is the purpose of this \emph{Letter} to show that, surprisingly, even in the absence of any coupling to the sterile neutrinos, such a light new vector boson can profoundly affect the population of the sterile neutrinos in the early universe through the mixing between left- and right-handed neutrinos, and it can yield the correct relic abundance of sterile neutrino DM.  

We consider an extension of the SM in which the $L_{\mu}-L_{\tau}$ current~\cite{He:1990pn,He:1991qd} is coupled to a new massive vector boson $\Zp$, with a mass in the range $\MZp \sim \MeV - \GeV$ and a coupling $\gZp \sim 10^{-6} - 10^{-2}$. This gauge interaction is well-hidden from most experiments due to its lack of coupling to first-generation leptons. The interaction can be described within a renormalizable field theory by gauging the $L_{\mu}-L_{\tau}$ current, and then breaking the gauge symmetry  with a new order parameter as in the electroweak theory (see for example~\cite{Altmannshofer:2014cfa} for a recent implementation)\footnote{Gauging other anomaly-free lepton currents, such as $B-L$, or $L$ with additional spectator fermions, give comparable sterile neutrino abundances to the $L_\mu-L_\tau$ theory, but with different $Z'$ phenomenology and considerably stronger  constraints.}. In addition, we add to the SM a sterile neutrino that mixes weakly with the active $\mu$- or $\tau$-neutrino states,
\be
\left(
\begin{array}{c}
 \nu_{\rm a}  \\
\nu_{\rm s}
\end{array}
\right)
\equiv 
\left(
\begin{array}{cc}
\cos\theta_0  &  \sin\theta_0   \\
-\sin\theta_0  & \cos\theta_0   
\end{array}
\right)
\left(
\begin{array}{c}
 \nu_1  \\
 \nu_2
\end{array}
\right),
\ee  
where $\nu_1$ and $\nu_2$ are the mass eigenstates, and $\nu_{\rm a}$ and $\nu_{\rm s}$ are the active and sterile ``flavour'' eigenstates, respectively. For concreteness in calculating the $\nu_{\rm s}$ abundance, we consider  the scenario in which the sterile neutrino  mixes exclusively with the flavour eigenstate $\nu_{\rm a}\equiv\nu_\tau$, although the extension to more general mixing is straightforward.

\begin{figure}[t]
\centering
\includegraphics[scale=0.8]{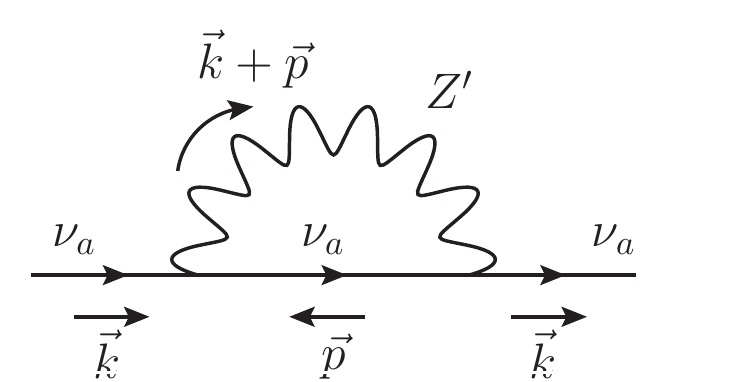}
\caption{Vector-boson contribution to the active-neutrino self-energy. Similar diagrams with SM vector bosons are also present.}
\label{fig:zprime_exchange}
\end{figure}

In the very early universe the vacuum mixing angle, $\theta_0$, is modified due to thermal corrections to active neutrino propagation in the hot medium \cite{Wolfenstein:1977ue}. The correction associated with the $Z'$ is depicted in Fig.~\ref{fig:zprime_exchange} with analogous diagrams for SM vector bosons. The mixing angle in medium is a function of both the temperature, $T$, as well as the neutrino momentum, $k$, and is given by~\cite{Dodelson:1993je}
\be
\label{eqn:mixing_in_medium}
\sin^2(2\theta_m) = \frac{\sin^2(2\theta_0)}{\sin^2(2\theta_0) + \left[ \cos(2\theta_0) + \DeltaR(k,T) \right]^2},
\ee 
where $\DeltaR$ is related to the real part of the contributions to the neutrino self-energy from vector-boson exchange, and is given by \cite{Notzold:1987ik,Wu:2009yr}
\be
\DeltaR(k,T) &\equiv& \DeltaR^{\rm (SM)}(k,T) + \DeltaR^{(Z')}(k,T). \\
\nonumber \\
\label{eqn:Delta_SM}
\DeltaR^{\rm (SM)}(k,T) &=& \frac{28\pi\sin^2\theta_{\rm W}\cos^2\theta_{\rm W} \,G_{\rm F}^2\, T^4 k^2}{45\alpha\,\mN^2},\label{eq:SM_finiteT}
\ee
$G_{\rm F}$ is the Fermi constant, and $\theta_{\rm W}$ is the Weinberg angle. In Eq.~(\ref{eq:SM_finiteT}), we show the contribution from SM $Z$ interactions in the limit $T\ll M_W,M_Z$; there also exists an analogous charged-current contribution \cite{Notzold:1987ik}, which depends on the active neutrino flavour eigenstate mixed with $\nu_{\rm s}$. The real part of the $\Zp$ contribution to the on-shell self-energy, $\DeltaR^{(Z')}(k,T)$, is determined by first computing the imaginary part $\Delta_{\rm I}^{(Z')}(k,T)$ of the diagram in Fig.~\ref{fig:zprime_exchange} in non-equilibrium thermal quantum field theory~\cite{Wu:2009yr},
\be
\nonumber
\DeltaI^{(Z')}(\omega) = \frac{\pi \gZp^2}{4} \int \frac{d^3p}{(2\pi)^3p E_{\vec{p}+\vec{k}} }\bigg( \Pi_0(\omega,\vec{p}) - \Pi_1(\omega,\vec{p}) \bigg), 
\ee
where the functions $\Pi_{0,1}(\omega,\vec{p})$ are given explicitly in Eqs.~(A.2) and~(A.3) of ref.~\cite{Wu:2009yr}. The real part, $\DeltaR^{(Z')}(k,T)$, then follows from the dispersion relation
\be\label{eq:dispersion}
\DeltaR^{(Z')}(k,T) = \frac{1}{\pi} \int_{-\infty}^{\infty} d\omega ~{\rm P.V.}\left( \frac{ \DeltaI^{(Z')}(\omega) }{\omega - k -i 0^+}\right),
\ee
where P.V.~specifies the principal value. We evaluate the integrals numerically without any assumptions about the relative magnitudes of $\MZp$, $\mN$, and $T$ \cite{Quimbay:1995jn,Wu:2009yr}. However, we do assume that the lepton asymmetry vanishes.

\begin{figure}[t]
\centering
\includegraphics[scale=0.7]{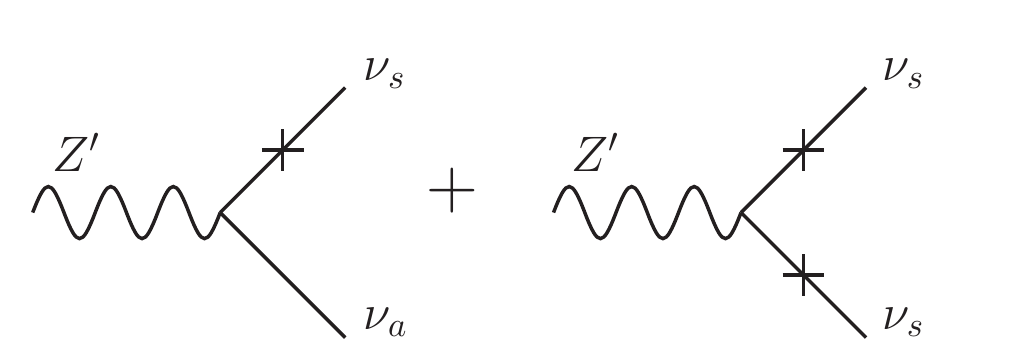}
\caption{Feynman diagrams corresponding to the lowest order processes that contribute to the sterile neutrino abundance through $Z'$ decay. }
\label{fig:zprime_decay}
\end{figure}

The self-energy corrections of keV-scale neutrinos from SM interactions alone, Eq.~(\ref{eqn:Delta_SM}), strongly suppress the mixing angle for $T\gtrsim150\,\,\mathrm{MeV}(\mN/\mathrm{keV})^{1/3}$~\cite{Dodelson:1993je}. This prevents any sizable production of sterile neutrinos at early epochs. As the Universe expands and cools below temperatures of a few hundred MeV, the active-sterile mixing is no longer strongly suppressed. 
Sterile neutrinos are then produced by $\Zp$ decay into active neutrino states and subsequent oscillation into $\nu_{\rm s}$, as in Fig.~\ref{fig:zprime_decay}. The rate for sterile neutrino production through this process is
\be
\Gamma_{Z' \rightarrow \nu_{\rm s}} = \frac{\gZp^2\MZp}{12\pi}\frac{ \sin^2(2\theta_{\rm m})}{4}\left(1 + \tan^2\theta_{\rm m}  \right),
\ee
where the second term in the bracket comes from the diagram on the right in Fig.~\ref{fig:zprime_decay} and is generally irrelevant except when the mixing angle is large, $\theta_{\rm m} \approx \pi/4$, as occurs when mixing is resonantly enhanced~\cite{Mikheyev:1989dy,Wu:2009yr}. This process is analogous to sterile neutrino production through $W$ and $Z$ decay~\cite{Wu:2009yr}, except that the abundance generated by these SM decays is too small to account for the DM abundance due to the large mixing angle suppression from thermal effects at $T\sim M_W,M_Z$. 

Sterile neutrinos are also produced in $2\rightarrow2$ scattering, with rate $\sim \gZp^4 T \sin^2(2\theta_{\rm m})\left(1 + \tan^2\theta_{\rm m}  \right)/4$, which is smaller than the rate of production from $\Zp$ decay by a factor of $\gZp^2 T/\MZp$.  For the small coupling regime we consider in this \emph{Letter}, this production mode is negligible compared to production from $\Zp$ decay at $T\sim \MZp$. If $T\gg \MZp$, $2\rightarrow2$ scattering can be important; however, the mixing angle in medium, $\theta_{\rm m}$ is suppressed in this epoch by thermal SM electroweak contributions according to Eq.~(\ref{eqn:mixing_in_medium}), and the sterile neutrino production rate is correspondingly minute. Therefore, the final sterile neutrino abundance comes predominantly from $\Zp$ decay.

The abundance of sterile neutrinos is governed by the Boltzmann equation, which is a reasonable approximation to the full quantum kinetic equations \cite{Abazajian:2001nj}. Using standard notation~\cite{Kolb:1990vq}, we define the yield as the number density of sterile neutrinos divided by the entropy density, $\YN \equiv n_{\rm s}/s$, and the inverse temperature variable $x \equiv \MZp/T$. In terms of these variables, the momentum-averaged Boltzmann equation takes the form \cite{Shaposhnikov:2006xi},
\be
\label{eqn:Boltzmann_Eq}
\frac{d\YN}{dx} &=&  \frac{45}{4\pi^4} \frac{x^4}{ g_{*\mathrm{S}}(x)M_{Z'}^2 H(M_{Z'})}\left(1-\frac{x}{4g_*} \frac{dg_{*}}{dx}\right) \\ \nonumber
&\times & \int_0^\infty dk\,\Gamma_{Z'\rightarrow \nu_{\rm s}}(x,k)\,\xi(x,k,M_{Z'}),
\ee
with
\be
\xi(x,k,\MZp) = \int_{k+M_{Z'}^2/4k}^\infty dE\,f_{Z'}(E,T),
\eea
where $H(T) = 1.66 \sqrt{\gstar} T^2/\Mpl$ is the Hubble rate during the radiation era (with $\Mpl = 1.22\times 10^{19}\GeV$ as the Planck scale), $\gstar(x)$ is the number of relativistic degrees of freedom during that epoch, $g_{*\mathrm{S}}$ is the contribution of relativistic degrees of freedom to the entropy density, and $f_{Z'}(E,T)$ is the equilibrium $\Zp$ distribution,
\be
f_{Z'}(E,T) = \frac{3}{e^{E/T}-1}.
\ee
The factor of three in the numerator corresponds to the three $\Zp$ polarization states. We keep track of the change in the number of relativistic degrees of freedom, $\gstar$, because the range of temperatures of interest to us has some overlap with the QCD phase transition, where $\gstar$ changes rapidly as a function of temperature.

The right-hand side of the Boltzmann equation~(\ref{eqn:Boltzmann_Eq}) depends in a complicated fashion on the temperature through $g_*$ and through the production rate $\Gamma_{\Zp\rightarrow \nu_{\rm s}}$, with the production rate's dependence on the temperature and neutrino momentum through the mixing angle in medium, Eq.~(\ref{eqn:mixing_in_medium}). In particular, the mixing angle may be resonantly enhanced if the denominator of Eq.~(\ref{eqn:mixing_in_medium}) vanishes \cite{Mikheyev:1989dy}. This can occur for certain neutrino momenta $k$ as a consequence of the thermal corrections to neutrino propagation, provided $\Delta_{\rm R}^{(Z')}$ is sufficiently large. A resonance occurs when the following is satisfied,
\be
\label{eqn:resonance_cond}
\cos2\theta_0 + \DeltaR \approx 1 + \Delta_{\rm R}^{\rm{(SM)}} + \Delta_{\rm R}^{(Z')} \approx 0.
\ee 
That such resonances are possible even in the absence of a lepton asymmetry was first realized in ref.~\cite{Wu:2009yr} in the context of SM vector-boson contributions to the mixing angle in the early universe. While, as we show below, the resonances do not ultimately contribute substantially to the $\nu_{\rm s}$ abundance they do warrant a careful consideration. We therefore discuss separately the regions of parameter space in which resonances do and do not occur.

\begin{figure}[t]
\centering
\includegraphics[width=0.43\textwidth]{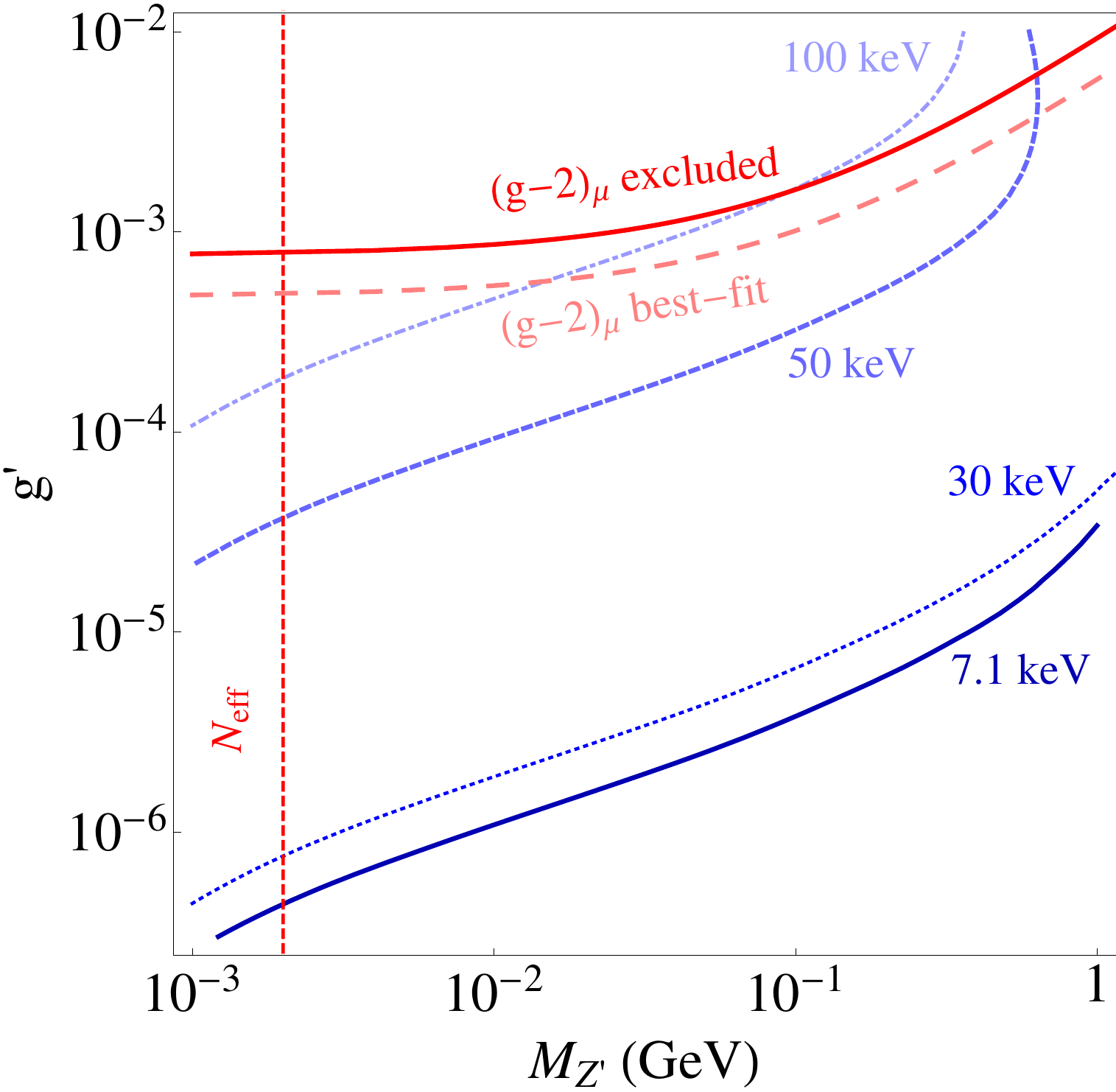}
\caption{A contour plot of the sterile neutrino abundance in the $\MZp - \gZp$ plane. Shown are the contours corresponding to the observed DM relic abundance, $Y_{\rm DM} = 4.7\times 10^{-4}({\rm keV}/\mN)$, for different values of the sterile neutrino mass $\mN$. The mixing angle  for each  $\mN$ is at the upper bound allowed by X-ray constraints \cite{Watson:2011dw}: $\sin2\theta_0=8\times10^{-6}$ ($\mN=7.1$ keV), $2.2\times10^{-6}$ ($\mN=30$ keV), $3.5\times10^{-8}$ ($\mN=50$ keV), $5\times10^{-9}$ ($\mN=100$ keV).  The solid-red curve depicts the $5\sigma$ bound muon $g-2$~\cite{Pospelov:2008zw}, and the dashed pink curve shows the coupling that best fits the muon $g-2$ anomaly. The vertical dashed curve corresponds to $\MZp \gtrsim 2.0\MeV$ coming from bounds on extra relativistic species in the early universe~\cite{Ade:2013zuv}.}
\label{fig:DM_abundance_curve}
\end{figure}

\noindent{\bf Non-resonant production:}
For sterile neutrinos in the range $\mN=1-100$ keV, a resonance requires $\gZp\gtrsim10^{-4}-10^{-3}$. Smaller values of the coupling give rise to exclusively non-resonant production of sterile neutrinos from $\Zp$ decay. The Boltzmann equation can then be solved numerically in a straightforward manner. The parametric dependence of the sterile neutrino abundance follows by dimensional analysis and power-counting of the coupling:
\be
\label{eqn:naive_yield}
Y_N \sim \left(\frac{45}{4\pi^4}\frac{1}{1.66 g_*^{3/2}}\right)  \frac{\gZp^2 M_{\rm Pl}}{\MZp}\sin^2(2\theta).
\ee
This relation is valid provided $g_*$ is not rapidly changing at $T\sim \MZp$.  When the mixing angle is approximately independent of $T$ and $\gZp$ near $T\sim \MZp$ (i.e.~for $\MZp\lesssim100$ MeV and $\gZp\lesssim10^{-4}-10^{-3}$), Eq.~(\ref{eqn:naive_yield}) exhibits a simple quadratic relationship between the coupling $\gZp$ and the mass $\MZp$ for fixed sterile neutrino abundance.

We solve the full Boltzmann equation numerically, determining for each $\Zp$ mass the necessary coupling to obtain the observed DM abundance. Our results are shown in Fig.~\ref{fig:DM_abundance_curve} for a few sterile neutrino masses and mixing angles. We choose mixing angles consistent with current upper bounds from X-ray observations \cite{Watson:2011dw}. Sterile neutrino masses of 30-100 keV are selected according to lower bounds from small-scale structure constraints, which range from $\mN\gtrsim8.8-22$ keV for a thermal sterile neutrino, depending on  assumptions made \cite{Abazajian:2005xn,Seljak:2006qw,Viel:2006kd,Boyarsky:2008xj,Polisensky:2010rw,deSouza:2013wsa,Viel:2013fqw,Horiuchi:2013noa}. We also consider $\mN = 7.1\keV$, whose radiative  decay  would give rise to an X-ray line around 3.55 keV, consistent with a recently observed excess in X-ray emissions at this energy from nearby galaxy clusters~\cite{Bulbul:2014sua,Boyarsky:2014jta}. We discuss the consistency of a 7.1 keV sterile neutrino with small-scale structure bounds shortly. 

The quadratic relationship between the $\Zp$ coupling and mass is apparent in Fig.~\ref{fig:DM_abundance_curve} for small coupling. Deviations from this relation emerge at larger coupling, when the thermal corrections from $\Zp$ increasingly suppress the mixing angle in medium. This suppression leads to the breakdown of the scaling in Eq.~(\ref{eqn:naive_yield}), and a still-larger coupling is required to obtain the correct DM abundance. The enhanced interaction rate associated with a larger coupling results in both the appearance of resonances in $\theta_{\rm m}$, as well as a quantum damping of active-sterile oscillations, as we discuss below in ``Resonant production''.

Also shown in Fig.~\ref{fig:DM_abundance_curve} are regions of the $\gZp-\MZp$ parameter space constrained by a variety of experiments and observations. These bounds depend on the specific $\Zp$ model, and we show them for the case of gauged $\mathrm{U}(1)_{\mu-\tau}$. If  $\MZp$ is too close to the temperature of  SM neutrino decoupling from the thermal bath, then $\Zp$ decays occurring after neutrino decoupling give a contribution to the number of effective neutrino species $(N_{\rm eff})$ in excess of the value allowed by Planck  at 95\% C.L.~\cite{Ade:2013zuv}; this gives the constraint $\MZp\gtrsim 2.0$ MeV. We also require that the $\Zp$-mediated decays of the sterile neutrino, $\nu_{\rm s}\rightarrow 2\nu+\bar\nu$ are sufficiently weak such that the sterile neutrino dark matter $\nu_{\rm s}$ has a cosmologically long lifetime. This  is  subdominant to other constraints for all sterile neutrino masses considered except $\mN=30$ keV, for which there is a bound of $\gZp\lesssim2.2\times10^{-3} \MZp/\mathrm{MeV}$.

Finally, the coupling $\gZp$ is  constrained by measurements of the muon anomalous magnetic dipole moment, and we show the $5\sigma$ bounds in Fig.~\ref{fig:DM_abundance_curve} as a conservative estimate \cite{Pospelov:2008zw}, since the current theoretical value is some $3\sigma$-discrepant compared with measurement \cite{Beringer:1900zz}. In fact, this long-standing discrepancy can be resolved with the same $\Zp$ responsible for sterile neutrino production. For example, in the regime where the $\Zp$ mass is well below $100\MeV$, its contribution to $(g-2)_\mu$ is approximately independent of its mass, and it accounts for the anomaly for $\gZp \approx 5\times10^{-4}$. Then the relation between $\Zp$ parameters accounting for both $(g-2)_\mu$ and the observed DM abundance is rather simple, 
\be
\MZp^{(g-2)_\mu} \sim  0.01\,\,\mathrm{GeV} \left(\frac{\sin2\theta_0}{5\times10^{-9}}\right)^2 \left(\frac{\mN}{100\,\,\mathrm{keV}}\right).
\ee
For a larger $\Zp$ mass, the precise parameters accounting for the $(g-2)_\mu$ anomaly are determined by solving the Boltzmann equation~(\ref{eqn:Boltzmann_Eq}) as  in Fig.~\ref{fig:DM_abundance_curve}.

\noindent{\bf Resonant production:} 
A resonance occurs when Eq.~(\ref{eqn:resonance_cond}) is satisfied. For temperatures below the weak scale, the SM contribution $\Delta_{\rm R}^{\rm (SM)}$ is strictly positive, and a resonance can only develop when $\Delta_{\rm R}^{(\Zp)} < 0$ and  $|\Delta_{\rm R}^{(\Zp)}|>1+\Delta_{\rm R}^{\rm (SM)}$. This can indeed take place for $T\sim \MZp$ and with couplings larger than $\gZp\gtrsim10^{-4}-10^{-3}$. For a given temperature $T$, there is then a critical momentum $k_{\rm c}(T)$ for which the resonance condition applies, and this momentum mode enjoys maximal mixing between active and sterile neutrinos $(\theta_{\rm m}=\pi/4$). This would na\"ively lead to a large production rate of sterile neutrinos of momentum $k_{\rm c}$ through $\Zp$ decays on-resonance.

However, the larger coupling in the resonance regime also implies stronger interaction of the active neutrino with the thermal bath. The decoherence from rapid scattering leads to a damping of  oscillations from active to sterile neutrinos, which is the well-known quantum Zeno effect~\cite{Misra:1976by,Stodolsky:1986dx,Abazajian:2001nj,Boyanovsky:2006it}. Whether the interaction rate is sufficiently large to suppress the oscillations is determined by the ratio of interaction rate to oscillation rate, which is related to the ratio of imaginary part of the self-energy to the real part \cite{Wu:2009yr},
\be\label{eq:damping}
\gamma(k,T) =  \frac{\DeltaI(k,T)}{\sqrt{\sin^2(2\theta_0) + \left[ \cos(2\theta_0) + \DeltaR(k,T) \right]^2 }}. 
\ee
Since the real and imaginary parts are closely related by Eq.~(\ref{eq:dispersion}), if $\gZp$ is large enough to satisfy the resonance condition ($\DeltaR \approx - \cos(2\theta_0)$), then damping is strong ($\gamma \approx \DeltaI/\sin(2\theta_0) \gg 1$). In this regime, the production of sterile neutrinos scales as
\be
\Gamma_{\Zp \rightarrow \nu_{\rm s}} =  \frac{ \sin^2(2\theta_m)}{\gamma^2} \Gamma_{\Zp \rightarrow \nu_{\rm a}}= \frac{ \sin^2(2\theta_0)}{\DeltaI^2}  \Gamma_{\Zp \rightarrow \nu_{\rm a}}.
\ee
Therefore, even though the mixing angle is maximal ($\theta_{\rm m}=\pi/4$), the damping of the oscillation exactly cancels the large enhancement in the mixing angle and leads to a production rate which is parametrically equal to the production rate of adjacent non-resonant momentum modes. In Fig.~\ref{fig:DM_abundance_curve} above, we included this effect whenever resonances and quantum damping are relevant ($\gZp\gtrsim10^{-4} -10^{-3}$). 
\vspace{1mm}

\noindent{\bf Momentum spectrum:} 
Since sterile neutrinos  can be cold, warm, or hot DM candidates, the $\nu_{\rm s}$ momentum spectrum is also important for determining constraints. In the case considered here of sterile neutrino production through $Z'$ decay, with no subsequent scattering of the sterile neutrinos, the $\nu_{\rm s}$ momentum distribution is non-thermal.  We find numerically that sterile neutrinos from $\Zp$ decay generally have $\langle k\rangle/T\lesssim2.4-2.5$ for $\MZp\lesssim1$ GeV (consistent with \cite{Shaposhnikov:2006xi}), in contrast with $\langle k\rangle/T \approx 3.15$ for a thermal distribution, and $\langle k \rangle/T\approx 2.8-2.9$ for production through the Dodelson-Widrow mechanism (SM weak interactions) \cite{Asaka:2006nq}.  The $\nu_{\rm s}$ spectrum is most important for $\mN$ close to the lower bounds from small-scale structure ($\mN\gtrsim8.8$ keV for Dodelson-Widrow production in the most conservative limit \cite{Horiuchi:2013noa}). For instance, the possible observation of an X-ray line from the decay of a 7.1 keV sterile neutrino \cite{Bulbul:2014sua,Boyarsky:2014jta} is in conflict with this bound unless a colder-than-thermal mechanism is responsible for sterile neutrino production. The decay of a $\Zp$  gives precisely such a spectrum, with an estimated free-streaming length of $\lambda_{\rm FS}\approx0.09\,\,\mathrm{Mpc}/h$ \cite{Bond:1980ha,Kolb:1990vq} that appears consistent with the 8.8 keV Dodelson-Widrow sterile neutrino bound \cite{Abazajian:2005gj}; however, a recent analysis suggests that 7.1 keV sterile neutrinos consistent with small-scale structure bounds may need to be much colder than predicted by our $\Zp$ model \cite{Abazajian:2014gza}. This merits further study, and we defer a detailed analysis to future work.

\begin{acknowledgments}

We thank Kev Abazajian, Eder Izaguirre, Marc Kamionkowski, Gordan Krnjaic, Kirill Melnikov, Josef Pradler, Maxim Pospelov, and Carlos Tamarit for helpful discussions. IY thanks the Johns Hopkins University for their hospitality. BS thanks the California Institute of Technology for their hospitality. BS is supported in part by the Canadian Institute of Particle Physics.  IY is supported in part by funds from the Natural Sciences and Engineering Research Council (NSERC) of Canada. This research was supported in part by Perimeter Institute for Theoretical Physics. Research at Perimeter Institute is supported by the Government of Canada through Industry Canada and by the Province of Ontario through the Ministry of Research and Innovation.
\end{acknowledgments}

\bibliography{sterile-bib}
\end{document}